\documentclass[twocolumn]{aastex63}

\def\rmit#1{{\it #1}}              
\def\specchar#1{{\sc #1}}

\def\NaIDtwo{\mbox{Na\,\specchar{i}\,\,D$_2$}}
\def\SiI{\mbox{Si\,\specchar{i}}}

\def\HeI{\mbox{He\,\specchar{i}}}

\def\eg{\rmit{e.g.}}

\newcommand{\GG}[1]{}

\received{2020 June 26}
\revised{2020 August 8}
\accepted{2020 August 24}

\shorttitle{Chromospheric resonant cavity in umbrae}
\shortauthors{Felipe et al.}

\graphicspath{{./}{figures/}}

\usepackage{multirow}

\begin{document}

\title{Chromospheric resonances above sunspots and potential seismological applications}

\correspondingauthor{Tobias Felipe}
\email{tobias@iac.es}

\author{Tobias Felipe}
\affiliation{Instituto de Astrof\'{\i}sica de Canarias \\
38205 C/ V\'{\i}a L{\'a}ctea, s/n, La Laguna, Tenerife, Spain}
\affiliation{Departamento de Astrof\'{\i}sica, Universidad de La Laguna \\
38205, La Laguna, Tenerife, Spain}

\author{Christoph Kuckein}
\affiliation{Leibniz-Institut f{\"u}r Astrophysik Potsdam (AIP) \\
An der Sternwarte 16, 14482 Potsdam, Germany}

\author{Sergio Javier Gonz\'alez Manrique}
\affiliation{Instituto de Astrof\'{\i}sica de Canarias \\
38205 C/ V\'{\i}a L{\'a}ctea, s/n, La Laguna, Tenerife, Spain}
\affiliation{Departamento de Astrof\'{\i}sica, Universidad de La Laguna \\
38205, La Laguna, Tenerife, Spain} 
\affiliation{Astronomical Institute, Slovak Academy of Sciences (AISAS) \\
05960 Tatransk\'{a} Lomnica, Slovak Republic} 

\author{Ivan Milic}
\affiliation{Department of Physics, University of Colorado, Boulder, CO 80309, USA}
\affiliation{Laboratory for Atmospheric and Space Physics, University of Colorado, Boulder, CO 80303, USA }
\affiliation{National Solar Observatory, Boulder, CO 80303, USA}

\author{C. R. Sangeetha}
\affiliation{Instituto de Astrof\'{\i}sica de Canarias \\
38205 C/ V\'{\i}a L{\'a}ctea, s/n, La Laguna, Tenerife, Spain}
\affiliation{Departamento de Astrof\'{\i}sica, Universidad de La Laguna \\
38205, La Laguna, Tenerife, Spain}

\begin{abstract}
Oscillations in sunspot umbrae exhibit remarkable differences between the photosphere and chromosphere. We evaluate two competing scenarios proposed for explaining those observations: a chromospheric resonant cavity and waves traveling from the photosphere to upper atmospheric layers. We have employed numerical simulations to analyze the oscillations in both models. They have been compared with observations in the low (\NaIDtwo) and high (\HeI\ 10830 \AA) chromosphere. The nodes of the resonant cavity can be detected as phase jumps or power dips, although the identification of the latter is not sufficient to claim the existence of resonances. In contrast, phase differences between velocity and temperature fluctuations reveal standing waves and unequivocally prove the presence of an acoustic resonator above umbrae. Our findings offer a new seismic method to probe active region chromospheres through the detection of resonant nodes.  

\end{abstract}


\section{Introduction} \label{sect:intro}

The nature of photospheric and chromospheric waves in active regions has intrigued solar physicists over several decades. Magnetic field concentrations are known to produce strong changes in the observed wavefield \citep{Braun+etal1987, Braun+etal1988}. The analysis of numerous observations \citep[\eg,][]{Beckers+Tallant1969, Giovanelli1972, Lites+etal1998, Bogdan+Judge2006}, in combination with theoretical modeling \citep[\eg,][]{Ferraro+Plumpton1958, Thomas1983, Cally+etal1994, Roberts2006} and numerical simulations\citep[\eg,][]{Bogdan+etal2003, Rosenthal+etal2002,Khomenko+Collados2006,Fedun+etal2011,Felipe+etal2011}, has led to the consensus that the observed oscillations are slow-magnetoacoustic waves travelling at the sound speed along magnetic field lines \citep[see][for a review]{Khomenko+Collados2015}. More uncertainties exist regarding the propagating or standing character of those waves and how this issue contributes to the observed change in the dominant period from five minutes at the photosphere to three minutes at the chromosphere.   

One of the proposed models is that of a chromospheric acoustic resonator \citep{Zhugzhda+Locans1981,Gurman+Leibacher1984,Zhugzhda2008}. In this scenario, the temperature gradients at the photosphere and transition region constitute the boundaries of a resonant cavity. However, phase difference spectra between Doppler velocities determined from photospheric and chromospheric spectral lines have shown indications of wave propagation \citep{Centeno+etal2006, Felipe+etal2010b, Cho+etal2015,Kanoh+etal2016, KrishnaPrasad+etal2017}. They support that waves with a period of three minutes observed at the chromosphere come directly from deeper photospheric layers through wave propagation. These high-frequency oscillations dominate the chromospheric signal due to the spatial attenuation of waves with frequencies below the cutoff value (approximately 5 mHz), since the latter form evanescent waves which do not propagate. Observational evidence of the photospheric excitation of waves with a period of three minutes by magnetoconvection has also been found through the detection of power enhancements of waves with three-minute period around umbral dots and light bridges \citep{Chae+etal2017}.

In the last years several works have pointed out the relevance of resonances in the umbral chromosphere \citep{Botha+etal2011, Snow+etal2015, Felipe2019} and photosphere \citep{Chae+etal2019}. Recently, \citet{Jess+etal2019} claimed the presence of a resonant cavity above a sunspot based on the detection of a high-frequency power peak in \HeI\ 10830 \AA\ observations. The use of this observable as an evidence of a resonant cavity was questioned by \citet{Felipe2020}, who showed that those power peaks are not commonly found in \HeI\ 10830 \AA\ data and that a similar power excess can be produced by non-linear effects, without the presence of a reflecting layer at the transition region. Here, we use numerical simulations to explore standing waves trapped in a resonant cavity and compare the results with observations. Resonances produce nodes, and at those locations a reduced oscillatory amplitude and phase jumps are expected. We have identified unambiguous measurements to discriminate between propagating and standing waves based on the analysis of the phase.

\section{Magnetohydrodynamic simulations} \label{sect:simulations}

\subsection{Numerical methods} \label{sect:methods}

The simulations presented in this work were developed with the code MANCHA3D \citep{Khomenko+Collados2006, Felipe+etal2010a}. We have computed wave propagation from below the photosphere to the corona in the umbral model M from \citet{Maltby+etal1986} with a 2000 G vertical magnetic field. We aim to study two cases: (i) oscillations of waves partially trapped in the chromosphere due to the gradients of the transition region, and (ii) waves that can freely propagate upward. For the first case, we have added a transition region to model M by imposing a temperature increase from the chromospheric temperature to an isothermal corona at 0.5 million Kelvin. We define the transition region as an atmospheric layer of a chosen thickness where the temperature gradient is imposed. In the second kind of simulations, we allow upward wave propagation by eliminating the transition region and maintaining an isothermal chromosphere from $z=1700$ km onwards. 

Simulations were performed in the 2.5D approximation, keeping vectors as three-dimensional objects but applying derivatives only in one vertical and one horizontal direction. The computational domain covers the vertical range from $z=-1140$ km to $z=3500$ km with a resolution of 10 km and $z=0$ defined at the height where the
optical depth at 5000 Å is unity. In the horizontal direction we set 96 points with a spatial step of 50 km. Waves are driven at $z=-150$ km by a perturbation in the vertical force derived from actual sunspot observations \citep{Felipe+etal2011, Felipe+Sangeetha2020}. 

In the following analyses, we employ the vertical velocity and temperature fluctuations obtained as outputs from the simulations. We study power and phase difference spectra. In the phase difference spectra between velocity at two different heights (V-V spectra), we use the sign convention to subtract the phase of the higher layer from the phase of the lower layer, so that a positive phase shift indicates upward wave propagation. In the phase spectra between velocity and temperature (V-T spectra), a positive phase difference indicates that the temperature fluctuations lag the velocity signal (with a positive velocity corresponding to redshifts).

\subsection{Chromospheric resonant cavity} \label{sect:resonance}

\begin{figure*}[ht!]
\plotone{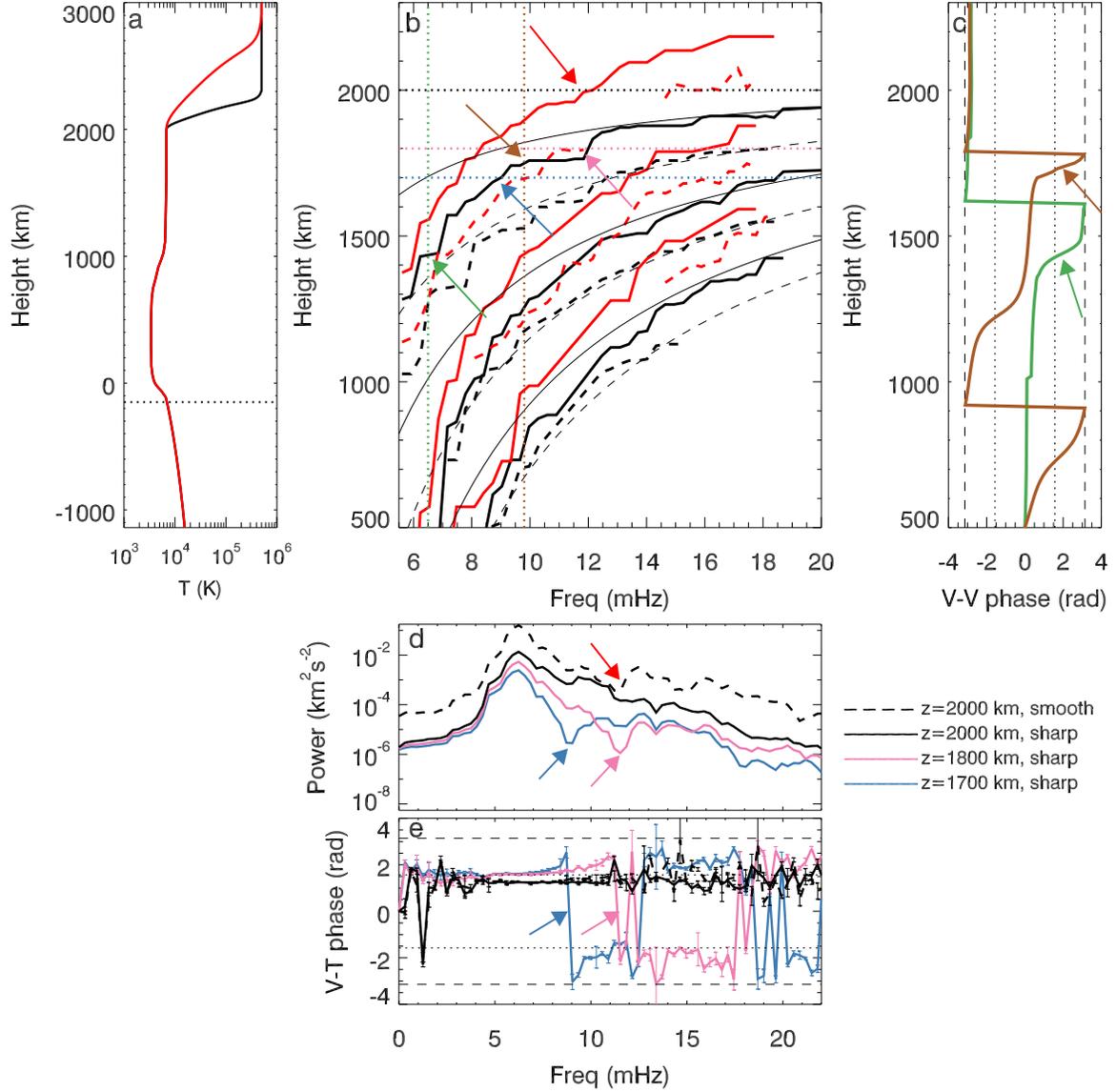}
\caption{Panel a: Temperature stratification of the background atmosphere for the simulations with a sharp (black) and smooth (red) transition region. The dotted line indicates the height of the driver. Panel b: Location of the velocity (thick solid lines) and temperature (thick dashed lines) nodal planes in the simulations with a sharp (black) and smooth (red) transition region. Thin solid (dashed) lines show the position of velocity (temperature) nodes in a model with a frequency-independent reflecting layer at $z=2050$ km and a wave speed of 9 km s$^{-1}$ \citep[\eg,][]{Fleck+Deubner1989}. Vertical (horizontal) dotted lines denote the frequencies (atmospheric heights) illustrated in panel c (d-e). Panel c: Vertical variation of the phase of the velocity oscillations for waves with a frequency of 6.5 (green) and 10.5 (brown) mHz. Panel d: Velocity power spectra at several heights, as given by the legend. Panel e: V-T spectra at the same heights. In panels d and e solid lines correspond to the simulation with a sharp transition region, whereas the dashed lines (hardly distinguishable from the black line in panel e) show the spectra from the simulation with a smooth temperature gradient. Arrows point to the position of some nodes of the standing wave in the Height-Frequency diagram (a) and their signatures in power (d) and phase (c, e) spectra. Error bars indicate the standard deviation. Straight dashed (dotted) lines in phase spectra indicate a phase shift of $\pm\pi$ ($\pm\pi/2$). 
\label{fig:resonance}}
\end{figure*}

The steep temperature gradients at the transition region partially reflect slow-magnetoacoustic waves, which are trapped between that atmospheric layer and the photospheric temperature gradients. Theory predicts the formation of a resonant cavity, which results in the presence of standing waves in the chromosphere. Figure \ref{fig:resonance} shows the results from two umbral simulations with the base of the transition region located at $z=2000$ km. One of them has a sharp temperature gradient at the transition region, whereas the other has a smoother gradient and a slightly higher amplitude was imposed at the driver (Figure \ref{fig:resonance}a). In the latter, we are interested in analyzing the effects of nonlinearities on the results.

The location of the velocity and temperature nodes in the simulations (Figure \ref{fig:resonance}b) was determined from the examination of the vertical variation of the oscillatory phases. They are identified as phase jumps of $\pi$ rad (see arrows in Figure \ref{fig:resonance}c). A comparison of the numerical nodal planes of the simulation with a sharp temperature gradient (thick black lines in Figure \ref{fig:resonance}b) with those determined from a simple model with a frequency-independent reflecting layer (thin black lines) shows a good match at high frequencies. In contrast, at lower frequencies velocity nodes are shifted to deeper layers, since low-frequency waves are reflected at a lower height in the transition region. Regarding the temperature nodes, their position is closer to the predicted by the model for all frequencies. This way, at low frequencies the location of the velocity and temperature nodes is very close. The case with a smoother transition region (red lines in Figure \ref{fig:resonance}b) exhibits a stronger variation of the nodal-plane height with frequency.

The chromospheric power spectra (Figure \ref{fig:resonance}d) show the expected maximum in the three-minute band ($\sim$6 mHz). In the following, we discuss how the nodes of the simulation with a sharp temperature gradient are revealed in phase and power spectra. At $z=1700$ km, the power exhibits a strong dip at the frequency where this atmospheric height intersects the location of the first nodal plane ($\sim$9 mHz, blue arrows in panels b and d). Another not so prominent dip is found at 18 mHz, where the second velocity node is at $z=1700$ km. For low frequencies, the phase spectra between velocity and temperature (Fig. \ref{fig:resonance}e) shows the $\pi/2$ rad phase difference expected from a standing mode \citep[\eg][]{Deubner1974,Al+etal1998}. At those frequencies where the selected geometrical height intersects a velocity or temperature nodal plane, a jump of $\pi$ rad is found. For the V-T spectra at $z=1700$ km, these phase jumps are seen at 9 mHz (velocity node), 12.5 mHz (temperature node), and 19 mHz (velocity node). Power and phase spectra obtained at different heights show a shift in the location of the features associated to nodes. For example, at $z=1800$ km the manifestation of the first velocity node (power dip and V-T phase jump, pink arrows) is shifted to a higher frequency (12 mHz) with respect to the spectra at $z=1700$ km, whereas the signature of the first temperature node (V-T phase jump) is found at $\sim$18 mHz. In the case of atmospheric heights above the first velocity nodal plane (\eg, $z=2000$ km), the power spectra lacks any nodal dip (Figure \ref{fig:resonance}d), and a $\pi/2$ rad phase difference is found in the V-T spectra for all frequencies (Figure \ref{fig:resonance}e).

\subsection{Standing and propagating oscillations} \label{sect:comparison}

\begin{figure}[ht!]
\plotone{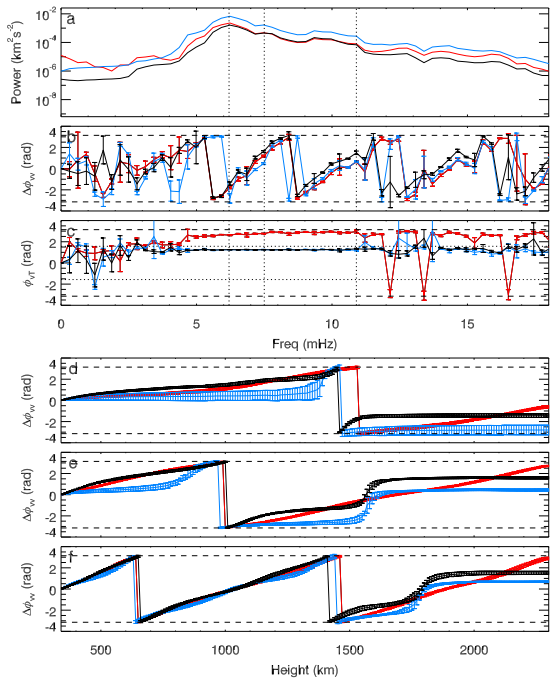}
\caption{Power and phase differences in a model without transition region (red) and in a resonant cavity in the adiabatic case (blue) and with radiative losses (black). The three panels of the top subset show the velocity power spectra at $z=2000$ km (a), V-V phase spectra between the signals at $z=340$ km and $z=2000$ km (b), and V-T phase spectra at $z=2000$ km (c). The three panels of the bottom subset show the vertical variation of the velocity phase for waves with a frequency of 6.2 (d), 7.5 (e), and 10.9 (f) mHz. Those frequencies are marked by vertical dotted lines in the top subset. Error bars indicate the standard deviation.    
\label{fig:standing_prop}}
\end{figure}

Our goal is to identify measurements that can be employed to discriminate between the standing waves produced by a resonant cavity and propagating waves. Figure \ref{fig:standing_prop} shows the oscillatory signatures of adiabatic waves and non-adiabatic waves \citep[radiative losses in the photosphere and low chromosphere implemented following Newton's cooling law with relaxation time given by][]{Spiegel1957} in models with a sharp transition region located at $z=2000$ km and those obtained from a model without transition region. The power spectra at $z=2000$ km \citep[approximately the formation height of the \HeI\ 10830 \AA\ triplet,][]{Avrett+etal1994} lack any feature that could potentially be used to distinguish between propagating and standing waves. In contrast, the simulation with a smooth temperature gradient exhibits at that height a small dip at the frequency where the velocity node intersects the velocity nodal plane (12 mHz, Figure \ref{fig:resonance}d).

For the upward propagating waves, the V-V spectra between photospheric and chromospheric velocities (Figure \ref{fig:standing_prop}b) shows the expected behavior, with a progressive increase in the phase with increasing frequency. In the frequency range between 4 and 9 mHz, the V-V spectrum of the simulation with adiabatic waves in a resonant cavity shows significant differences, since the phase difference varies in sudden $\pi$ rad increments, instead of a gentle increase. However, if we take into account radiative losses, the V-V spectrum from the resonant cavity is very similar to that measured for purely propagating waves. 

The bottom subset of Figure \ref{fig:standing_prop} illustrates the wave propagation at some selected frequencies. The phase of propagating waves monotonically increases with height (the plotted frequencies are above the cutoff value). The blue line shows a clear standing pattern, with $\pi$ rad jumps at the locations of the velocity nodes. At the photosphere and low chromosphere, the phase of the waves in the resonant cavity with radiative losses is similar to those of the simulation with no upper wave reflections, but the standing pattern is revealed by the presence of nodes at higher chromospheric layers. All the atmosphere above the first velocity node oscillates in phase.   

Figure \ref{fig:standing_prop}c shows the V-T spectra at $z=2000$ km of the simulations discussed in this section. For propagating waves we obtain the $\pi$ rad phase difference predicted by theory, whereas all simulations with transition region exhibit a $\pi/2$ phase shift. The simulation with a smooth temperature gradient does not exhibit any phase jump in the V-T spectra at $z=2000$ km (Figure \ref{fig:resonance}e), even though a velocity nodal plane crosses that geometrical height. In this case, there is a thin layer of a few hundred kilometers where the node is only visible in the power spectra. However, a simulation with the same background atmosphere but using a very low amplitude driver (keeping the simulations in the linear regime) actually exhibits the phase jump at the frequency of the node. Thus, the absence of the phase jump in V-T spectra is related to non-linear effects.

\section{Oscillatory signatures in solar observations} \label{sect:observations}
Observations were taken at the Vacuum Tower Telescope \citep[VTT,][]{vonderLuhe1998} and at the GREGOR telescope \citep{Schmidt+etal2012}. At VTT we used the Tenerife Infrared Polarimeter
\citep[TIP,][]{MartinezPillet+etal1999, Collados+etal2007}. At GREGOR, data was taken with the GREGOR Infrared Spectrograph \citep[GRIS,][]{Collados+etal2012} and the
GREGOR Fabry-P\'erot Interferometer \citep[GFPI,][]{Puschmann+etal2012}. 
A summary of all observations is listed in Table \ref{tab:observations}.

\begin{table*}[!htb]
\begin{center}
\caption{Details of the six data sets obtained with VTT and GREGOR telescopes}\label{tab:observations}
\begin{tabular}{cllllllll}
\hline
\hline
Instrument/Telescope    & AR         & Date & Spectral line & Duration & Cadence  & Figure & Reference \rule[-4pt]{0pt}{15pt}  \\
                        &  NOAA      &      & (\AA)         & (s) &     (s) &  &  \rule[-4pt]{0pt}{15pt}  \\
\hline
\hline

TIP/VTT                 & 09173  & 2000 Oct 01   & \HeI\ 10830          & 3555     & 7.9      & \ref{fig:HeI10830} a-b  & \citet{Centeno+etal2006}    \rule[-4pt]{0pt}{14pt} \\

TIP/VTT                 & 09443  & 2001 May 09   & \HeI\ 10830          & 4200     & 2.1      & \ref{fig:HeI10830} c-d  & \citet{Centeno+etal2006}\\

TIP/VTT                 & 10969  & 2007 Aug 29   & \HeI\ 10830          & 4482     & 18       & \ref{fig:HeI10830} e-f  & \citet{Felipe+etal2010b}\\

GRIS/GREGOR             & 12662  & 2017 Jun 17   & \HeI\ 10830          & 4603       & 5.7       & \ref{fig:HeI10830} g-h  & \citet{Felipe+etal2018b}\\

\multirow{2}{7.5em}{GRIS/GREGOR}  & \multirow{2}{5em}{12662}  & \multirow{2}{6em}{2017 Jun 18}   & \multirow{2}{5em}{\HeI\ 10830}          & \multirow{2}{4em}{2217}       & \multirow{2}{4em}{5.6}       &\multirow{2}{4em}{\ref{fig:HeI10830} i-j}  & Same setup from \\
 &   &    &          &        &        &   & \citet{Felipe+etal2018b}\\

GFPI/GREGOR             & 12708  & 2018 May 09   & \NaIDtwo\ 5889        & 1422      & 31.6       & \ref{fig:NaD2}           & \citet{Kuckein2019}\\
\hline
\end{tabular}
\end{center}
\end{table*}

\subsection{High-chromosphere phase spectra: \HeI 10830 \AA} \label{sect:HeI10830}

The simulations presented in the previous section point to the V-T phase spectra as the best measurement to discriminate between propagating and standing waves. We have performed a Milne-Eddington inversion \citep{Socas-Navarro2001} in the umbral region of five \HeI\ 10830 \AA\ temporal series. From the inversions we extracted the Doppler velocity and width of the spectral line. The latter is proportional to the square root of the temperature \citep{delToroIniesta2003, Borrero+etal2014}. We have employed it to derive the phase of temperature fluctuations. 

\begin{figure*}[ht!]
\plotone{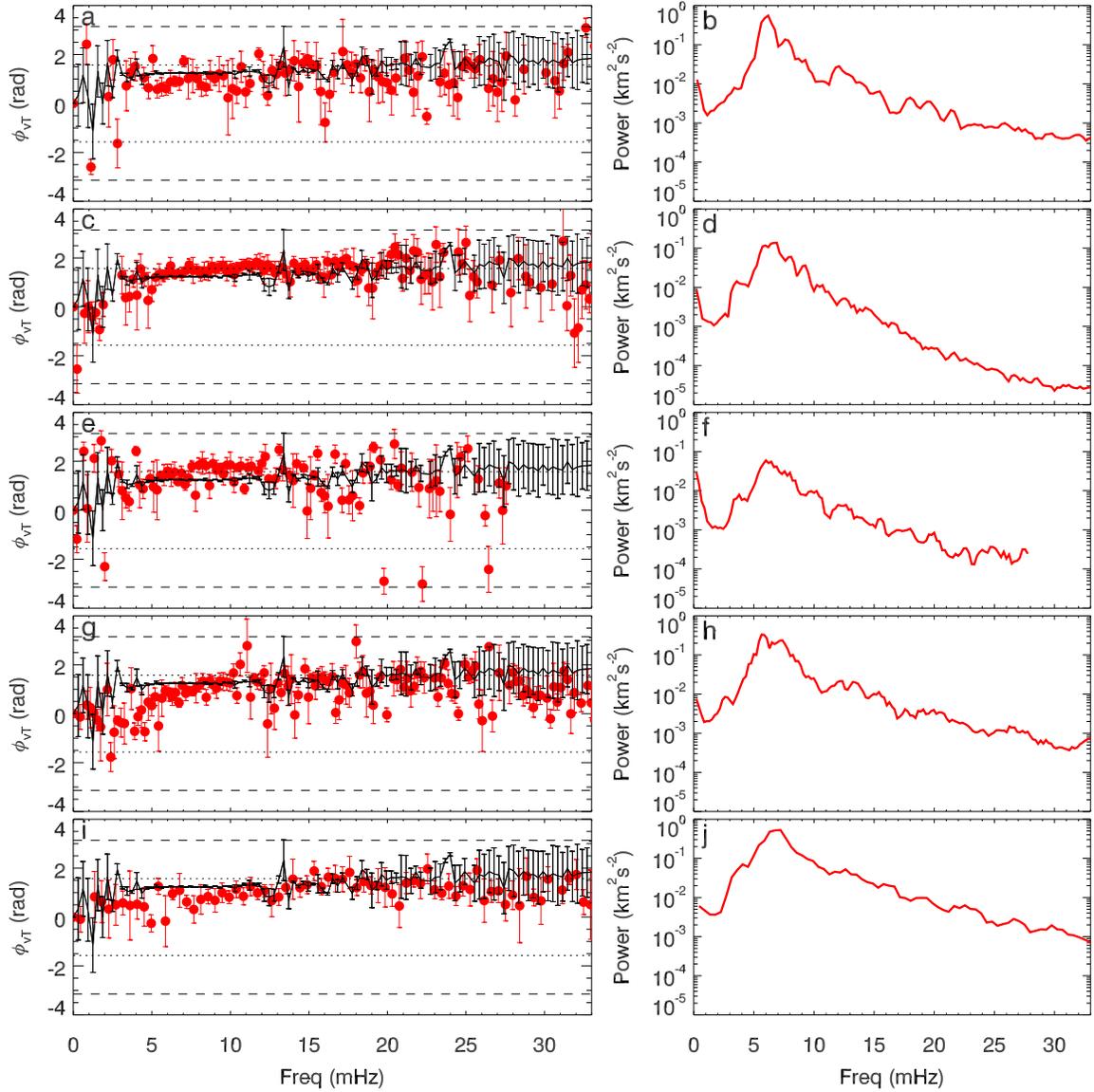}
\caption{Phase and power spectra measured with the \HeI\ 10830 \AA\ line for five different umbral observations. Left panels: Observed (red circles) and simulated (with resonant cavity, black lines) V-T spectra. Error bars show the standard deviation. Right panels: Velocity power spectra. See Table \ref{tab:observations} for a summary of the data.
\label{fig:HeI10830}}
\end{figure*}

The umbral \HeI\ 10830 \AA\ observations show an approximately constant V-T phase difference around $\pi/2$ for frequencies beyond $\sim 4$ mHz (Figure \ref{fig:HeI10830}). This result is consistent with the presence of standing modes. The absence of phase jumps in all the probed frequencies indicates that the line is formed above the first velocity nodal plane or at the thin layer at the base of a smooth transition region where V-T spectra do not show jumps at the frequency of the node. We have compared the V-T spectra with the values predicted by a simulation with a resonant cavity and assumed the formation height of the \HeI\ 10830 \AA\ line at $z=2000$ km. They show a remarkable agreement. The discrepancies between model and observations are comparable to the differences between observational sets.

\subsection{Low-chromosphere phase spectra: \NaIDtwo} \label{sect:NaD2}

\begin{figure}[ht!]
\plotone{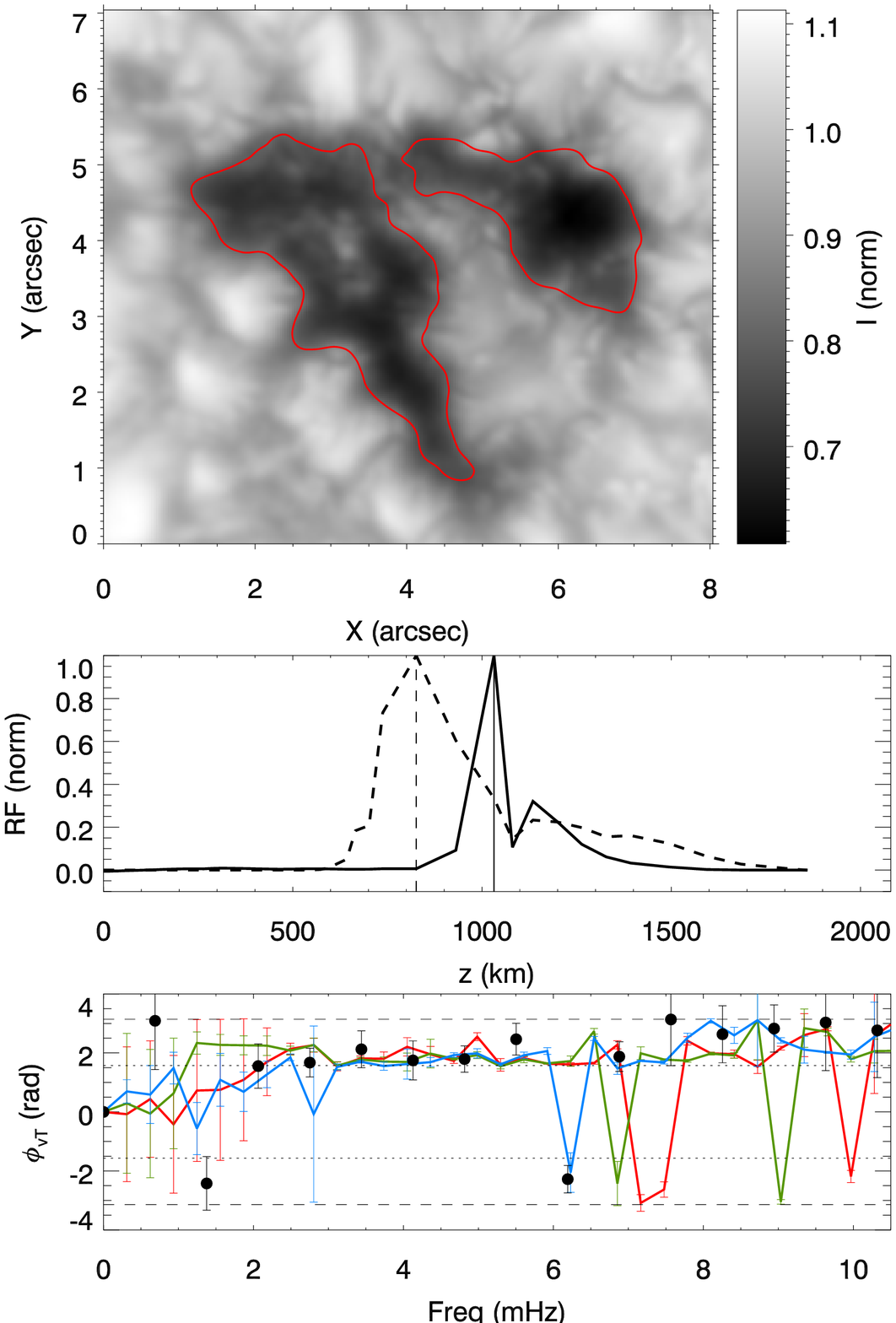}
\caption{Observations of a pore in NOAA 12708. Top panel: broad-band image. Middle panel: Response functions of the \NaIDtwo\ intensity in umbral model M \citep{Maltby+etal1986} for temperature (solid line) and velocity (dashed line) at 110 m\AA\ from the line center. Vertical lines indicate the geometrical height of the temperature ($z=1030$ km, solid) and velocity ($z=830$ km, dashed) signals used for the numerical V-T spectrum. Bottom panel: V-T spectrum from the \NaIDtwo\ line averaged in the region delimited by the red contours in the top panel (black circles) and from numerical simulations with the base of the transition region at $z=2200$ km (red), $z=2400$ km (green), and $z=2600$ km (blue). Error bars show the standard deviation. \label{fig:NaD2}}
\end{figure}

The velocity and intensity of the \NaIDtwo\ line where measured from bisectors taken at 10\% of the line profile, with 0\% corresponding to the core. Intensity fluctuations have been assumed as representative of temperature oscillations. The V-T spectrum of the \NaIDtwo\ (Figure \ref{fig:NaD2}) also shows a $\pi/2$ rad phase difference, confirming the presence of standing modes at the low chromosphere. However, at 6.2 mHz it exhibits a $-\pi/2$ rad phase shift. This phase difference change is caused by two successive phase jumps, the first one produced by the intersection of the line formation height with a velocity nodal plane followed by another jump due to the intersection of a temperature node. At the low chromosphere, velocity and temperature nodal planes are located at very close frequencies (Figure \ref{fig:resonance}b). In a phase spectra, they produce two consecutive $\pi$ rad phase jumps over a small frequency range. 

We have compared the observed \NaIDtwo\ V-T spectrum with those obtained from numerical simulations with the transition region at different heights. The formation height of temperature and velocity fluctuations of the \NaIDtwo\ line in the umbral atmosphere was determined from the computation of the response functions \citep[Figure \ref{fig:NaD2}b,][]{Milic+vanNoort2018}. The simulation with the transition region at $z=2600$ km exhibits a better agreement with the observed phase spectrum since it captures the phase jump at 6.2 mHz. An upward shift of the transition region produces an upward displacement of the nodal planes. According to Figure \ref{fig:resonance}b, this means that phase spectra probed at a fixed geometrical height will exhibit the $\pi$ rad phase jumps at lower frequencies as the transition region is displaced to higher atmospheric layers. This allows us to determine the height of the transition region from the identification of the nodes in the phase spectra.

\section{Discussion} \label{sect:conclusions}

We have proved the presence of a chromospheric resonant cavity above active regions from the comparison of spectroscopic observations with numerical modelling. From the numerical simulations, we have identified phase spectra between velocity and temperature (intensity) as a prominent measurement to discriminate between the standing waves of a resonant cavity and propagating waves. The evaluation of these phase spectra in actual observations reveals that standing waves are taking place in active region chromospheres. 

The analysis of V-T spectra to address the nature of the observed waves is a common approach from many pioneering studies of solar atmospheric waves \citep[\eg,][]{Schmieder1976, Lites+etal1982,Staiger+etal1984,Fleck+Deubner1989, Deubner+etal1990}. Here, we have obtained the phase relations from the analysis of numerical simulations. Our results are consistent with theoretical estimates. In addition, they offer more flexibility to model the resonant structure of the solar atmosphere. They take into account the frequency-dependent reflecting layer at the transition region \citep[the height where waves are reflected, given by the cutoff frequency, is poorly described by analytical expressions,][]{Felipe+Sangeetha2020}. Interestingly, the simulations show that the distance between velocity and temperature nodal planes depends on the frequency and the atmospheric height. This fact has implications for the interpretation of the observed standing waves in the low-chromosphere V-T spectra, which show a $-\pi/2$ rad phase difference (instead of $\pi/2$ rad) over a small frequency range. This phase jump has been observed and employed to locate the nodal planes and derive the height of the transition region. This example illustrates the new seismic analyses that can be performed following our findings. Since the location of the nodal planes depends mainly on the height and temperature gradients of the transition region and the wave speed, the identification of the nodes in the V-T spectra can be employed to derive the properties of the transition region and the chromospheric sound speed. 

Our results settle several puzzling results from wave studies in sunspots. The progressive phase increase measured in V-V spectra between photospheric and chromospheric lines, instead of the phase jumps expected from standing modes, has been interpreted as an evidence of the propagating nature of waves in those atmospheric layers \citep[\eg,][]{Centeno+etal2006, Kanoh+etal2016}. We have shown that this propagation only takes place between the photosphere and low chromosphere thanks to the effect of the radiative losses (Figure \ref{fig:standing_prop}). The assumption of propagating waves has been previously employed to derive the formation height of the \HeI\ 10830 \AA\ triplet in sunspots, obtaining $z\sim 1400$ km \citep[around 1000 km above the formation height of the \SiI\ 10827 \AA\ line,][]{Centeno+etal2006, Felipe+etal2010b}. This is in contrast with the formation height of $z\sim 2000$ km inferred from the modeling of the spectral line \citep{Avrett+etal1994}. We found that velocity oscillations are in phase above the first node (Figure \ref{fig:resonance}c) and, thus, this method is insensitive to variations in the formation height above that height.

Our results directly address the recent claims from \citet{Jess+etal2019}, who reported the presence of a chromospheric cavity based on the identification of a high-frequency power peak in \HeI\ 10830 \AA\ observations. Here, we have showed that the resonant structure of the sunspot atmosphere can manifest in the power spectra as dips (power enhancements) at the frequencies where the formation height of the line coincides with nodes (anti-nodes). No power dip/excess is found above the height of the first nodal plane. This result has not been previously reported in simulations of chromospheric resonances \citep{Botha+etal2011, Snow+etal2015}, but an examination of their results shows that they are consistent with our conclusions. For example, the power spectra presented by \citet{Botha+etal2011} (with the transition region at $z\sim 2000$ km) exhibit a dip at $z=1500$ km, but not at $z=2000$ km. Accommodating our findings with the numerical conclusions from \citet{Jess+etal2019} is more troubling. They found a high-frequency power peak in simulations with the transition region at different heights, including cases with the temperature gradient well below the selected formation of the \HeI\ 10830 \AA\ line. They cannot be understood as the imprint of the nodes/anti-nodes of the resonant cavity. In addition, they do not report frequency shifts in the location of the power peaks associated to displacements of the transition region, which would be the most evident manifestation of the resonances on the power spectra. More studies evaluating the effect of the simulation parameters in the power spectra are required. These analyses will improve by synthesizing and interpreting chromospheric lines \citep[\eg,][]{Felipe+etal2018b} instead of extracting the signals at a specific geometrical height, since the response function of the lines generally samples a broad range of atmospheric heights.

In this work, we have confirmed that in most cases the power spectra of the \HeI\ 10830 \AA\ line do not exhibit such strong high-frequency peaks. \citet{Jess+etal2020} argued that this peak is only visible under ideal observational conditions. The upcoming data from the next generation of solar telescopes will clarify whether this peak is unusual or a common feature hidden in most observations up to date. Our observations show some small dips in the \HeI\ 10830 \AA\ power spectra. We consider two possible interpretations. On the one hand, the frequency of that power excess ($\sim$12 mHz) agrees with the expected location of the harmonic of the main power peak ($\sim$6 mHz), as suggested by \citet{Felipe2020} and found in observational analyses \citep{Chae+etal2018}. On the other hand, they can be produced by the intersection of the formation height of the line with a velocity nodal plane under conditions similar to those represented by our simulation with a smooth transition region, since they are not associated to phase jumps in the V-T spectra (Figure \ref{fig:HeI10830}). We cannot discard that the power peak reported by \citet{Jess+etal2019} is actually a manifestation of the chromospheric cavity. However, we stress that such claim cannot be done with the simple examination of the power spectra.

\acknowledgments

Financial support from the State Research Agency (AEI) of the Spanish Ministry of Science, Innovation and Universities (MCIU) and the European Regional Development Fund (FEDER) under grant with reference PGC2018-097611-A-I00 is gratefully acknowledged. Funding from the H2020 projects SOLARNET (824135) and ESCAPE (824064) is gratefully acknowledged by CK. SJGM acknowledges the support of grants PGC2018-095832-B-I00 (MCIU), ERC-2017-CoG771310-PI2FA (European Research Council), and VEGA2/0048/20 (Slovak Academy of Sciences). We acknowledge the contribution of Teide High-Performance Computing facilities to the results of this research. TeideHPC facilities are provided by the Instituto Tecnol\'ogico y de Energ\'ias Renovables (ITER, SA). URL: \url{http://teidehpc.iter.es}. The 1.5-meter GREGOR solar telescope was built by a German consortium under the leadership of the Leibniz-Institut f\"ur Sonnenphysik in Freiburg (KIS) with the Leibniz-Institut f\"ur Astrophysik Potsdam (AIP), the Institut f\"ur Astrophysik G\"ottingen (IAG), the Max-Planck-Institut f\"ur Sonnensystemforschung in G\"ottingen (MPS), and the Instituto de Astrof\'isica de Canarias (IAC), and with contributions by the Astronomical Institute of the Academy of Sciences of the Czech Republic (ASCR). We thank H. Balthasar for his help during the GFPI observations. 

%

\vspace{5mm}
\facilities{GREGOR(GRIS and GFPI), VTT(TIP)}











\begin{thebibliography}{}
\expandafter\ifx\csname natexlab\endcsname\relax\def\natexlab#1{#1}\fi
\providecommand{\url}[1]{\href{#1}{#1}}
\providecommand{\dodoi}[1]{doi:~\href{http://doi.org/#1}{\nolinkurl{#1}}}
\providecommand{\doeprint}[1]{\href{http://ascl.net/#1}{\nolinkurl{http://ascl.net/#1}}}
\providecommand{\doarXiv}[1]{\href{https://arxiv.org/abs/#1}{\nolinkurl{https://arxiv.org/abs/#1}}}

\bibitem[{{Al} {et~al.}(1998){Al}, {Bendlin}, \& {Kneer}}]{Al+etal1998}
{Al}, N., {Bendlin}, C., \& {Kneer}, F. 1998, \aap, 336, 743

\bibitem[{{Avrett} {et~al.}(1994){Avrett}, {Fontenla}, \&
  {Loeser}}]{Avrett+etal1994}
{Avrett}, E.~H., {Fontenla}, J.~M., \& {Loeser}, R. 1994, in IAU Symposium,
  Vol. 154, Infrared Solar Physics, ed. D.~M. {Rabin}, J.~T. {Jefferies}, \&
  C.~{Lindsey}, 35

\bibitem[{{Beckers} \& {Tallant}(1969)}]{Beckers+Tallant1969}
{Beckers}, J.~M., \& {Tallant}, P.~E. 1969, \solphys, 7, 351

\bibitem[{{Bogdan} \& {Judge}(2006)}]{Bogdan+Judge2006}
{Bogdan}, T.~J., \& {Judge}, P.~G. 2006, Royal Society of London Philosophical
  Transactions Series A, 364, 313

\bibitem[{{Bogdan} {et~al.}(2003){Bogdan}, {Carlsson}, {Hansteen}, {McMurry},
  {Rosenthal}, {Johnson}, {Petty-Powell}, {Zita}, {Stein}, {McIntosh}, \&
  {Nordlund}}]{Bogdan+etal2003}
{Bogdan}, T.~J., {Carlsson}, M., {Hansteen}, V.~H., {et~al.} 2003, \apj, 599,
  626, \dodoi{10.1086/378512}

\bibitem[{{Borrero} {et~al.}(2014){Borrero}, {Lites}, {Lagg}, {Rezaei}, \&
  {Rempel}}]{Borrero+etal2014}
{Borrero}, J.~M., {Lites}, B.~W., {Lagg}, A., {Rezaei}, R., \& {Rempel}, M.
  2014, \aap, 572, A54, \dodoi{10.1051/0004-6361/201424584}

\bibitem[{{Botha} {et~al.}(2011){Botha}, {Arber}, {Nakariakov}, \&
  {Zhugzhda}}]{Botha+etal2011}
{Botha}, G.~J.~J., {Arber}, T.~D., {Nakariakov}, V.~M., \& {Zhugzhda}, Y.~D.
  2011, \apj, 728, 84, \dodoi{10.1088/0004-637X/728/2/84}

\bibitem[{{Braun} {et~al.}(1987){Braun}, {Duvall}, \&
  {Labonte}}]{Braun+etal1987}
{Braun}, D.~C., {Duvall}, Jr., T.~L., \& {Labonte}, B.~J. 1987, \apjl, 319,
  L27, \dodoi{10.1086/184949}

\bibitem[{{Braun} {et~al.}(1988){Braun}, {Duvall}, \&
  {Labonte}}]{Braun+etal1988}
---. 1988, \apj, 335, 1015, \dodoi{10.1086/166988}

\bibitem[{{Cally} {et~al.}(1994){Cally}, {Bogdan}, \&
  {Zweibel}}]{Cally+etal1994}
{Cally}, P.~S., {Bogdan}, T.~J., \& {Zweibel}, E.~G. 1994, \apj, 437, 505,
  \dodoi{10.1086/175014}

\bibitem[{{Centeno} {et~al.}(2006){Centeno}, {Collados}, \& {Trujillo
  Bueno}}]{Centeno+etal2006}
{Centeno}, R., {Collados}, M., \& {Trujillo Bueno}, J. 2006, \apj, 640, 1153,
  \dodoi{10.1086/500185}

\bibitem[{{Chae} {et~al.}(2018){Chae}, {Cho}, {Song}, \&
  {Litvinenko}}]{Chae+etal2018}
{Chae}, J., {Cho}, K., {Song}, D., \& {Litvinenko}, Y.~E. 2018, \apj, 854, 127,
  \dodoi{10.3847/1538-4357/aaa8e2}

\bibitem[{{Chae} {et~al.}(2019){Chae}, {Kang}, \& {Litvinenko}}]{Chae+etal2019}
{Chae}, J., {Kang}, J., \& {Litvinenko}, Y.~E. 2019, \apj, 883, 72,
  \dodoi{10.3847/1538-4357/ab3d2d}

\bibitem[{{Chae} {et~al.}(2017){Chae}, {Lee}, {Cho}, {Song}, {Cho}, \&
  {Yurchyshyn}}]{Chae+etal2017}
{Chae}, J., {Lee}, J., {Cho}, K., {et~al.} 2017, \apj, 836, 18,
  \dodoi{10.3847/1538-4357/836/1/18}

\bibitem[{{Cho} {et~al.}(2015){Cho}, {Bong}, {Nakariakov}, {Lim}, {Park},
  {Chae}, {Yang}, {Park}, \& {Yurchyshyn}}]{Cho+etal2015}
{Cho}, K.~S., {Bong}, S.~C., {Nakariakov}, V.~M., {et~al.} 2015, \apj, 802, 45,
  \dodoi{10.1088/0004-637X/802/1/45}

\bibitem[{{Collados} {et~al.}(2007){Collados}, {Lagg}, {D{\'{\i}}az
  Garc{\'{\i}} A}, {Hern{\'a}ndez Su{\'a}rez}, {L{\'o}pez L{\'o}pez}, {P{\'a}ez
  Ma{\~n}{\'a}}, \& {Solanki}}]{Collados+etal2007}
{Collados}, M., {Lagg}, A., {D{\'{\i}}az Garc{\'{\i}} A}, J.~J., {et~al.} 2007,
  in Astronomical Society of the Pacific Conference Series, Vol. 368, The
  Physics of Chromospheric Plasmas, ed. P.~{Heinzel}, I.~{Dorotovi{\v c}}, \&
  R.~J. {Rutten}, 611

\bibitem[{{Collados} {et~al.}(2012){Collados}, {L{\'o}pez}, {P{\'a}ez},
  {Hern{\'a}ndez}, {Reyes}, {Calcines}, {Ballesteros}, {D{\'{\i}}az}, {Denker},
  {Lagg}, {Schlichenmaier}, {Schmidt}, {Solanki}, {Strassmeier}, {von der
  L{\"u}he}, \& {Volkmer}}]{Collados+etal2012}
{Collados}, M., {L{\'o}pez}, R., {P{\'a}ez}, E., {et~al.} 2012, AN, 333, 872,
  \dodoi{10.1002/asna.201211738}

\bibitem[{{del Toro Iniesta}(2003)}]{delToroIniesta2003}
{del Toro Iniesta}, J.~C. 2003, Introduction to Spectropolarimetry (Cambridge:
  Cambridge University Press)

\bibitem[{{Deubner}(1974)}]{Deubner1974}
{Deubner}, F.-L. 1974, \solphys, 39, 31, \dodoi{10.1007/BF00154969}

\bibitem[{{Deubner} {et~al.}(1990){Deubner}, {Fleck}, {Marmolino}, \&
  {Severino}}]{Deubner+etal1990}
{Deubner}, F.~L., {Fleck}, B., {Marmolino}, C., \& {Severino}, G. 1990, \aap,
  236, 509

\bibitem[{{Fedun} {et~al.}(2011){Fedun}, {Shelyag}, \&
  {Erd{\'e}lyi}}]{Fedun+etal2011}
{Fedun}, V., {Shelyag}, S., \& {Erd{\'e}lyi}, R. 2011, \apj, 727, 17,
  \dodoi{10.1088/0004-637X/727/1/17}

\bibitem[{{Felipe}(2019)}]{Felipe2019}
{Felipe}, T. 2019, \aap, 627, A169, \dodoi{10.1051/0004-6361/201935784}

\bibitem[{{Felipe}(2020)}]{Felipe2020}
---. 2020, Nature Astronomy, \dodoi{10.1038/s41550-020-1157-5}

\bibitem[{{Felipe} {et~al.}(2010{\natexlab{a}}){Felipe}, {Khomenko}, \&
  {Collados}}]{Felipe+etal2010a}
{Felipe}, T., {Khomenko}, E., \& {Collados}, M. 2010{\natexlab{a}}, \apj, 719,
  357, \dodoi{10.1088/0004-637X/719/1/357}

\bibitem[{{Felipe} {et~al.}(2011){Felipe}, {Khomenko}, \&
  {Collados}}]{Felipe+etal2011}
---. 2011, \apj, 735, 65, \dodoi{10.1088/0004-637X/735/1/65}

\bibitem[{{Felipe} {et~al.}(2010{\natexlab{b}}){Felipe}, {Khomenko},
  {Collados}, \& {Beck}}]{Felipe+etal2010b}
{Felipe}, T., {Khomenko}, E., {Collados}, M., \& {Beck}, C. 2010{\natexlab{b}},
  \apj, 722, 131, \dodoi{10.1088/0004-637X/722/1/131}

\bibitem[{{Felipe} {et~al.}(2018){Felipe}, {Kuckein}, \&
  {Thaler}}]{Felipe+etal2018b}
{Felipe}, T., {Kuckein}, C., \& {Thaler}, I. 2018, \aap, 617, A39,
  \dodoi{10.1051/0004-6361/201833155}

\bibitem[{{Felipe} \& {Sangeetha}(2020)}]{Felipe+Sangeetha2020}
{Felipe}, T., \& {Sangeetha}, C.~R. 2020, \aap, 640, A4,
  \dodoi{10.1051/0004-6361/202038387}

\bibitem[{{Ferraro} \& {Plumpton}(1958)}]{Ferraro+Plumpton1958}
{Ferraro}, C.~A., \& {Plumpton}, C. 1958, \apj, 127, 459

\bibitem[{{Fleck} \& {Deubner}(1989)}]{Fleck+Deubner1989}
{Fleck}, B., \& {Deubner}, F.-L. 1989, \aap, 224, 245

\bibitem[{{Giovanelli}(1972)}]{Giovanelli1972}
{Giovanelli}, R.~G. 1972, \solphys, 27, 71

\bibitem[{Gurman \& Leibacher(1984)}]{Gurman+Leibacher1984}
Gurman, J.~B., \& Leibacher, J.~W. 1984, \apj, 283, 859

\bibitem[{{Jess} {et~al.}(2020{\natexlab{a}}){Jess}, {Snow}, {Houston},
  {Botha}, {Fleck}, {Krishna Prasad}, {Asensio Ramos}, {Morton}, {Keys},
  {Jafarzadeh}, {Stangalini}, {Grant}, \& {Christian}}]{Jess+etal2019}
{Jess}, D.~B., {Snow}, B., {Houston}, S.~J., {et~al.} 2020{\natexlab{a}},
  Nature Astronomy, 4, 220, \dodoi{10.1038/s41550-019-0945-2}


\bibitem[{{Jess} {et~al.}(2020{\natexlab{b}}){Jess}, {Snow}, {Fleck},
  {Stangalini}, \& {Jafarzadeh}}]{Jess+etal2020}
{Jess}, D.~B., {Snow}, B., {Fleck}, B., {Stangalini}, M., \& {Jafarzadeh}, S.
  2020{\natexlab{b}}, Nature Astronomy, \dodoi{10.1038/s41550-020-1158-4}


\bibitem[{{Kanoh} {et~al.}(2016){Kanoh}, {Shimizu}, \&
  {Imada}}]{Kanoh+etal2016}
{Kanoh}, R., {Shimizu}, T., \& {Imada}, S. 2016, \apj, 831, 24,
  \dodoi{10.3847/0004-637X/831/1/24}

\bibitem[{{Khomenko} \& {Collados}(2006)}]{Khomenko+Collados2006}
{Khomenko}, E., \& {Collados}, M. 2006, \apj, 653, 739, \dodoi{10.1086/507760}

\bibitem[{{Khomenko} \& {Collados}(2015)}]{Khomenko+Collados2015}
---. 2015, Living Reviews in Solar Physics, 12, 6, \dodoi{10.1007/lrsp-2015-6}

\bibitem[{{Krishna Prasad} {et~al.}(2017){Krishna Prasad}, {Jess}, {Van
  Doorsselaere}, {Verth}, {Morton}, {Fedun}, {Erd{\'e}lyi}, \&
  {Christian}}]{KrishnaPrasad+etal2017}
{Krishna Prasad}, S., {Jess}, D.~B., {Van Doorsselaere}, T., {et~al.} 2017,
  \apj, 847, 5, \dodoi{10.3847/1538-4357/aa86b5}

\bibitem[{{Kuckein}(2019)}]{Kuckein2019}
{Kuckein}, C. 2019, \aap, 630, A139, \dodoi{10.1051/0004-6361/201935856}

\bibitem[{{Lites} {et~al.}(1998){Lites}, {Thomas}, {Bogdan}, \&
  {Cally}}]{Lites+etal1998}
{Lites}, B.~W., {Thomas}, J.~H., {Bogdan}, T.~J., \& {Cally}, P.~S. 1998, \apj,
  497, 464, \dodoi{10.1086/305451}

\bibitem[{{Lites} {et~al.}(1982){Lites}, {White}, \&
  {Packman}}]{Lites+etal1982}
{Lites}, B.~W., {White}, O.~R., \& {Packman}, D. 1982, \apj, 253, 386,
  \dodoi{10.1086/159642}

\bibitem[{{Maltby} {et~al.}(1986){Maltby}, {Avrett}, {Carlsson},
  {Kjeldseth-Moe}, {Kurucz}, \& {Loeser}}]{Maltby+etal1986}
{Maltby}, P., {Avrett}, E.~H., {Carlsson}, M., {et~al.} 1986, \apj, 306, 284,
  \dodoi{10.1086/164342}

\bibitem[{{Martinez Pillet} {et~al.}(1999){Martinez Pillet}, {Collados},
  {S{\'a}nchez Almeida}, {Gonz{\'a}lez}, {Cruz-Lopez}, {Manescau}, {Joven},
  {Paez}, {Diaz}, {Feeney}, {S{\'a}nchez}, {Scharmer}, \&
  {Soltau}}]{MartinezPillet+etal1999}
{Martinez Pillet}, V., {Collados}, M., {S{\'a}nchez Almeida}, J., {et~al.}
  1999, in Astronomical Society of the Pacific Conference Series, Vol. 183,
  High Resolution Solar Physics: Theory, Observations, and Techniques, ed.
  T.~R. {Rimmele}, K.~S. {Balasubramaniam}, \& R.~R. {Radick}, 264

\bibitem[{{Mili{\'c}} \& {van Noort}(2018)}]{Milic+vanNoort2018}
{Mili{\'c}}, I., \& {van Noort}, M. 2018, \aap, 617, A24,
  \dodoi{10.1051/0004-6361/201833382}

\bibitem[{{Puschmann} {et~al.}(2012){Puschmann}, {Denker}, {Kneer}, {Al
  Erdogan}, {Balthasar}, {Bauer}, {Beck}, {Bello Gonz{\'a}lez}, {Collados},
  {Hahn}, {Hirzberger}, {Hofmann}, {Louis}, {Nicklas}, {Okunev},
  {Mart{\'{\i}}nez Pillet}, {Popow}, {Seelemann}, {Volkmer}, {Wittmann}, \&
  {Woche}}]{Puschmann+etal2012}
{Puschmann}, K.~G., {Denker}, C., {Kneer}, F., {et~al.} 2012, Astronomische
  Nachrichten, 333, 880, \dodoi{10.1002/asna.201211734}

\bibitem[{{Roberts}(2006)}]{Roberts2006}
{Roberts}, B. 2006, Philosophical Transactions of the Royal Society of London
  Series A, 364, 447, \dodoi{10.1098/rsta.2005.1709}

\bibitem[{{Rosenthal} {et~al.}(2002){Rosenthal}, {Bogdan}, {Carlsson}, {Dorch},
  {Hansteen}, {McIntosh}, {McMurry}, {Nordlund}, \&
  {Stein}}]{Rosenthal+etal2002}
{Rosenthal}, C.~S., {Bogdan}, T.~J., {Carlsson}, M., {et~al.} 2002, \apj, 564,
  508, \dodoi{10.1086/324214}

\bibitem[{{Schmidt} {et~al.}(2012){Schmidt}, {von der L{\"u}he}, {Volkmer},
  {Denker}, {Solanki}, {Balthasar}, {Bello Gonzalez}, {Berkefeld}, {Collados},
  {Fischer}, {Halbgewachs}, {Heidecke}, {Hofmann}, {Kneer}, {Lagg}, {Nicklas},
  {Popow}, {Puschmann}, {Schmidt}, {Sigwarth}, {Sobotka}, {Soltau}, {Staude},
  {Strassmeier}, \& {Waldmann }}]{Schmidt+etal2012}
{Schmidt}, W., {von der L{\"u}he}, O., {Volkmer}, R., {et~al.} 2012,
  Astronomische Nachrichten, 333, 796, \dodoi{10.1002/asna.201211725}

\bibitem[{{Schmieder}(1976)}]{Schmieder1976}
{Schmieder}, B. 1976, \solphys, 47, 435, \dodoi{10.1007/BF00154756}

\bibitem[{{Snow} {et~al.}(2015){Snow}, {Botha}, \&
  {R{\'e}gnier}}]{Snow+etal2015}
{Snow}, B., {Botha}, G.~J.~J., \& {R{\'e}gnier}, S. 2015, \aap, 580, A107,
  \dodoi{10.1051/0004-6361/201526115}

\bibitem[{{Socas-Navarro}(2001)}]{Socas-Navarro2001}
{Socas-Navarro}, H. 2001, in Astronomical Society of the Pacific Conference
  Series, Vol. 236, Advanced Solar Polarimetry -- Theory, Observation, and
  Instrumentation, ed. M.~{Sigwarth}, 487

\bibitem[{Spiegel(1957)}]{Spiegel1957}
Spiegel, E.~A. 1957, \apj, 126, 202

\bibitem[{{Staiger} {et~al.}(1984){Staiger}, {Mattig}, {Schmieder}, \&
  {Deubner}}]{Staiger+etal1984}
{Staiger}, J., {Mattig}, W., {Schmieder}, B., \& {Deubner}, F.~L. 1984,
  \memsai, 55, 147

\bibitem[{{Thomas}(1983)}]{Thomas1983}
{Thomas}, J.~H. 1983, Annual Review of Fluid Mechanics, 15, 321,
  \dodoi{10.1146/annurev.fl.15.010183.001541}

\bibitem[{{von der L{\"u}he}(1998)}]{vonderLuhe1998}
{von der L{\"u}he}, O. 1998, {New Astron. Rev.}, 42, 493,
  \dodoi{10.1016/S1387-6473(98)00060-8}

\bibitem[{{Zhugzhda}(2008)}]{Zhugzhda2008}
{Zhugzhda}, Y.~D. 2008, \solphys, 251, 501, \dodoi{10.1007/s11207-008-9251-3}

\bibitem[{{Zhugzhda} \& {Locans}(1981)}]{Zhugzhda+Locans1981}
{Zhugzhda}, Y.~D., \& {Locans}, V. 1981, Soviet Astronomy Letters, 7, 25

\end{thebibliography}
\end{document}